\documentclass[twocolumn,letter]{jpsj3}

\title{Interplay of Superconductivity, Antiferromagnetism, and Pauli Depairing in CeCoIn$_5$ }

\author{Carley \textsc{Paulsen}\thanks{E-mail: carley.paulsen@grenoble.cnrs.fr}, Dai \textsc{Aoki}$^{1}$, Georg \textsc{Knebel}$^{1}$, and Jacques \textsc{Flouquet}$^{1}$\thanks{E-mail: jacques.flouquet@cea.fr}} 

\inst{Institut N\'{e}el, MCBT Department, CNRS and Universit\'{e} Joseph Fourier, BP 166, 38042 Grenoble Cedex 9, France \\
$^{1}$SPSMS, UMR-E CEA / UJF-Grenoble 1, INAC, Grenoble, F-38054, France
}
\abst{Low temperature magnetization $M$ measurements down to 70~mK have been performed through the superconducting upper critical field $H_{c2}$ of CeCoIn$_5$ with the field $H$ parallel to the $c$ axis of the tetragonal crystal. The crossing through the upper critical field $H_{c2} (0)$ is marked by a sign change in the temperature dependence of $M(T)$. A Fermi liquid law of the magnetization is observed in the superconducting phase while in the paramagnetic state above $H_{c2}$ a quasi-linear temperature dependence is obeyed. The specific heat below $H_{c2}$  is therefor dominated by the increase of $\gamma (H)$ , the linear term of the specific heat, and is marked by a singularity at $H_{c2} (0)$. No indication of the so called $Q$-phase was observed below $H_{c2}$(in contrast to the case with $H$ applied in the basal plane).  }

\kword{CeCoIn$_5$, upper critical field, quantum criticality, magnetization measurements}

\begin{document}
\maketitle

Investigations of the heavy fermion Ce-115 family has lead to spectacular discoveries of new phases with a strong interplay between magnetism and superconductivity (SC) as shown in CeRhIn$_5$ with the re-entrance of antiferromagnetism (AF) inside the superconducting phase under high pressure and magnetic field.\cite{Park2006, Knebel2006}

CeCoIn$_5$ is even a more fascinating case than CeRhIn$_5$ as it was established by neutron \cite{Kenzelmann2008} and NMR \cite{Young2007, Koutroulakis2010} experiments that for a field in the basal plane an AF phase referred as the $Q$-phase appears in the high field low temperature (HFLT) boundary pegged to $H_{c2}(0)$, although above $H_{c2}(0)$ the ground state is paramagnetic (PM). Prior to this recent experimental evidence, it was suggested \cite{Radovan2003, Bianchi2003b, Matsuda2007}  that the Fulde-Ferrel Larkin-Ovchinnikov (FFL0) state \cite{Fulde1964, Larkin1964}  occurs in the HFLT region, with finite momentum pairing created by the strength of the Pauli limit. Recent theoretical studies emphasize the key role of the Pauli depairing of the Cooper pairs \cite{Ikeda2010b} and the possibility of a staggered field induced triplet component associated with the appearance of an AF phase coupled with or without a FFLO state \cite{Yanase2009, Aperis2010}

Here we report on detailed magnetization $M(T,H)$ experiments with the field $H$ aligned along the $c$ axis of the tetragonal crystal in order to search where are the singularities in the field dependence of the Sommerfeld coefficient $\gamma$ of the specific heat and to determine if signatures of such a $Q$-phase are also present for this direction. For $H$ along the $c$ axis it was also suggested that $H_{c2}(0)$ corresponds to an AF quantum critical point \cite{Paglione2003, Bianchi2003c}. 
We have  focused on the  SC region close to $H_{c2}(0)$ for $H \parallel c$ by using small field steps in order to look for evidence of the $Q$-phase. A HFLT phase has been reported to occur along the $H\parallel c$ axis when $H > 4.7$~T and below $T \sim 0.5$~K based on NMR and $\mu$SR experiments \cite{Oyaizu2007, Spehling2009}. However, no distinct transition line has been determined.
The vortex lattice change from  rhombic to  hexagonal  \cite{Bianchi2008} close to $H_{c2} (0)$ at $H_{hex} \approx 4.4$~T is directly linked to the Pauli limit \cite{Bianchi2008, Michal2010}.

Single crystals of CeCoIn$_5$ were obtained by the In-flux method. Crystals grown previously in the same way in our laboratory have been used for NMR \cite{Koutroulakis2010} and resistivity \cite{Howald2011} measurements. 

\begin{figure}
\begin{center}
\includegraphics[width=0.75\hsize,clip]{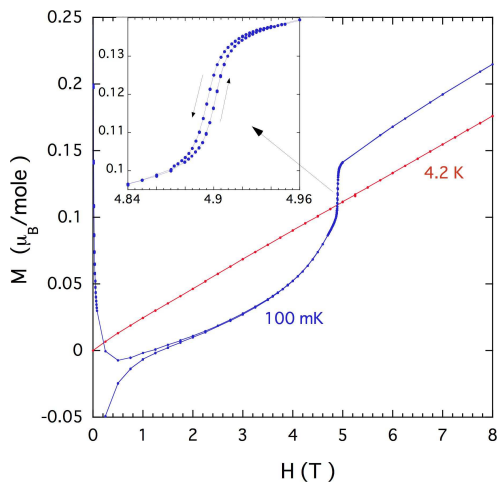}
\end{center}
\caption{(Color online) Magnetization curve $M(H)$ of CeCoIn$_5$ for $H \parallel c$ at $T = 4.2$~K and $T = 100$~mK, the insert shows a zoom close to $H_{c2} (0)$. }
\label{f1}
\end{figure}

\begin{figure}
\begin{center}
\includegraphics[width=0.75\hsize,clip]{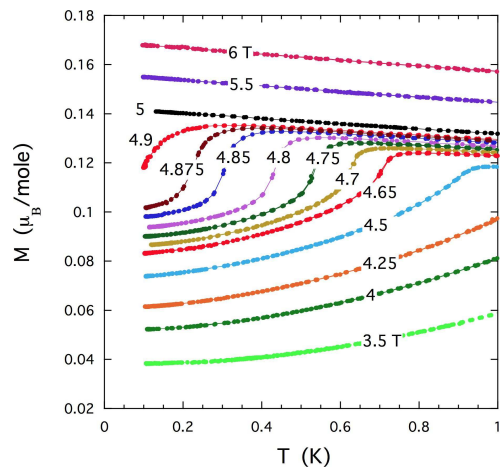}
\end{center}
\caption{(Color online) Temperature variation of the magnetization at constant magnetic field $H \parallel c$ through $H_{c2}(0)$ in a linear representation.  }
\label{f2}
\end{figure}

\begin{figure}
\begin{center}
\includegraphics[width=0.75\hsize,clip]{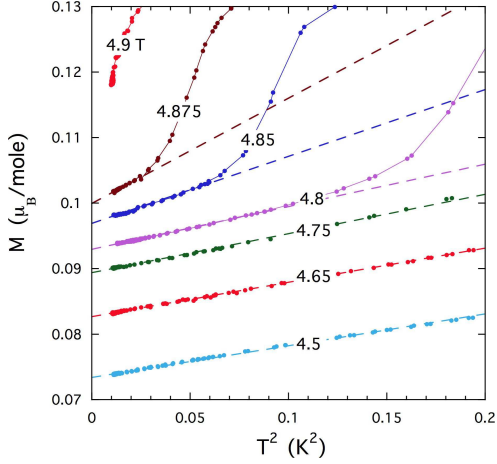}
\end{center}
\caption{(Color online) Magnetization $M$ as function of $T^2$ for different magnetic fields $H<H_{c2}(0)$ applied parallel to the $c$ axis. Dashed lines indicate the a fit with a $\beta (H)T^2$ dependence. }
\label{f3}
\end{figure}

The magnetization measurements were performed with a low temperature-high field SQUID magnetometer developed at Institut N\'{e}el. A miniature dilution refrigerator allows for absolute value measurements by the extraction method at temperatures down to $T=70$~mK.  Notwithstanding, it is well known that large peak effects are present in this sample, and thus even a slight inhomogeneity of the magnetic field, can give a large spurious signal when using the extraction method, as seen in a difference between up and down extraction measurements. For our experimental setup, consistent  results are obtained by averaging the up and down measurements together. 
We digress to elucidate this subtle but important experimental detail: At $H =2$~T,  CeCoIn$_5$ is near the maximum of the peak effect, and the field gradient in our magnetometer is approximately 1~Gauss over our normal scan length of 35~mm. At this field, we found that the difference between up and down extractions was  approximately 0.007~$\mu_B$/mole, or nearly the  same magnitude as the averaged magnetization. (However, this difference becomes much smaller above 3.5~T where it is less than 10\% of the value of the magnetization, and drops to zero near  4.8~T.)   During measurements of $M$ vs $T$ at fixed $H$, we have used two techniques in order to verify that averaging the up and down extractions is correct for this sample: One method is to reduce the scan length of the extraction. As the scan length is decreased, the difference between up and down extractions becomes smaller, and for scans below 15 mm, we could eliminate completely the difference, with the results converging directly on top of the averaged normal scan length data, although with more scatter in the data points. A second method is to place the sample in one of the detection coils of the magnetometer, and then to vary the temperature without moving the sample at all, and thus measure the relative difference in magnetization at fixed field as a function of temperature. (This technique was used for previous studies of the pseudo-metamagnetism of CeRu$_2$Si$_2$ \cite{Paulsen1990}.) Again, these results (when correctly offset) fall on the averaged up and down extraction data. This gives us confidence in our method, and all the data shown in Figs.\ref{f1},\ref{f2}, and \ref{f3} are the averaged up and down measurements.

Figure \ref{f1} shows $M$ vs. $H$ up to 8~T taken at $T=4.2$~K and at $T = 100$~mK, well above and below the superconducting transition at $T_c = 2.3$~K in zero field.  Good agreement is found with previous published magnetization data \cite{Tayama2002},  although  the hysteresis and the observed peak effects in this sample are significantly smaller than previously reported. This is certainly due to the difference between the field sweep technique of Ref.\citen{Tayama2002}  vs.~the extraction method used here. But in addition, a part may be due to different sample qualities and their preparation.
The value of  $H_{c2}(0)=4.96$~T  is in very good agreement with other bulk measurements like the specific heat \cite{Bianchi2003b}, thermal expansion \cite{Zaum2010} and NMR \cite{Oyaizu2007}. The hysteresis at $H_{c2}(0)$ is quite small, $50 \times 10^{-4}$~T and the jump of the magnetization is $\Delta M \approx 3.5 \times 10^{-2} \mu _B$, from an initial value of 0.1~$\mu _B$ just below $H_{c2}(0)$. If the SC transition would be second order down to $T=0$, a straightforward extrapolation of the magnetization data suggests that $H_{c2}(0)$ would be near 7 T.

Our main interest in $M(T)$ at constant field presented in this letter is that it is related to the field dependence of the specific heat coefficient $C/T$ via the Maxwell relation 
\begin{equation*} 
\frac{\partial (C/T)}{\partial H} = \frac{\partial ^2 M}{\partial T^2}.
\end{equation*}
The high accuracy of the measurements associated with a careful determination of the reversible component makes possible an excellent estimate of $\partial (C/T)/\partial H$. In addition, because it is a field derivative method of the specific heat, it is quite sensitive not only to phase transitions but even to field induced crossovers. The Maxwell relation allows to obtain the bulk macroscopic magnetic field evolution of the specific heat, of course without the separation between the vortex and the supercurrent contributions.

If the magnetization, assumed to be of purely electronic origin, follows a characteristic Fermi-liquid temperature dependence $M(T) \approx \beta (H)T^2$, the coefficient $\beta (H)$ is directly linked to the field derivative of the Sommerfeld coefficient $\partial \gamma /\partial H$ of the linear specific heat $\gamma (H)$. 

Figure \ref{f2} shows the temperature dependence of $M(T)$ at constant field above and below $H_{c2}(0)$. The first observation is that for fields $H < H_{c2}(0)$ the magnetization decreases on lowering the temperature while for $H > H_{c2}(0)$, $M(T)$ increases on cooling. Clearly, the sign change in $\partial M(H)/\partial T$ marks a singularity at $H_{c2}$. 
Another important observation is illustrated in Fig. \ref{f3} where $M$ is plotted vs. $T^2$: for $H < H_{c2}(0)$ and for low enough temperatures, $M(T)$ is seen to follow a $T^2$ dependence. However, above $H_{c2}(0)$  a dominant linear temperature dependence of $M (T)$ has been observed down to the lowest measured temperature $T=70$~mK. This last observation is still consistent with previous experiments where a Fermi-liquid regime was abserved above $H_{c2}(0)$, but only at very low temperature $T<40$~mK \cite{Howald2011}. The clear observation of a $T^2$ dependence of $M(H)$ for $H < H_{c2} (0)$ demonstrates that the field dependence of the specific heat is governed by the growth of the linear term $\gamma (H)$ with magnetic field and not by other contributions. Similar effects have also been reported recently for $H\ll H_{c2} (0)$ when $H$ is applied in the basal plane \cite{An2010}. The huge field increase of $\gamma$ with field is associated with the gapless tendency of the SC on approaching $H_{c2}$ (see Refs.\citen{Brandt1967, Dukan1997, Vavilov1998}) and the large values of $\gamma$ of the heavy fermion quasi-particles in their normal phase close to the magnetic quantum critical point.

\begin{figure}
\begin{center}
\includegraphics[width=0.75\hsize,clip]{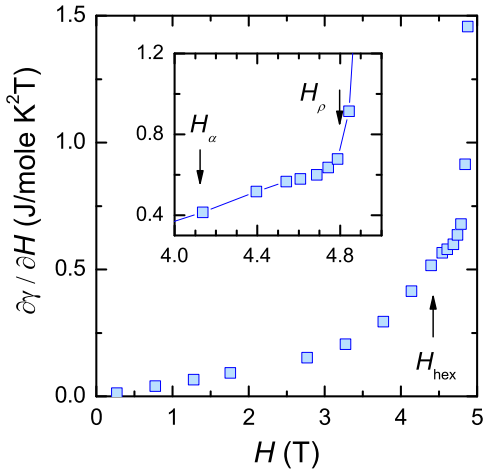}
\end{center}
\caption{(Color online) Field dependence of $\partial \gamma /\partial H$ below $H_{c2}(0)$. The arrow indicates the position of the vortex lattice phase transition \cite{Bianchi2008} at $H_{hex} = 4.4$~T. In the insert we indicate $H_\alpha \approx 4.1$~T and $H_{\rho} \approx 4.8$ derived in Refs.~\citen{Zaum2010, Howald2011}.    }
\label{dgammadH}
\end{figure}

\begin{figure}
\begin{center}
\includegraphics[width=0.8\hsize,clip]{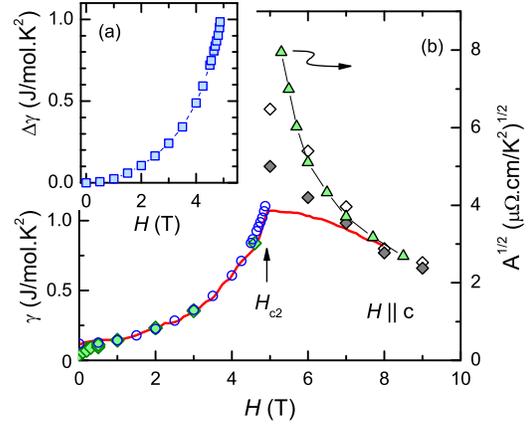}
\end{center}
\caption{(Color online) Inset (a): Relative field dependence of $\Delta \gamma (H)$ in the SC phase derived from magnetization measurements as explained in the text. Main frame (b): Comparison of $\gamma (H)$ $vs. H$ determined by different experiments. Solid line is $\gamma (H) vs. H$ at $T = 250$~mK from Ref.~\citen{Ikeda2001}, solid diamonds from Ref.~\citen{Bianchi2003c} for $H>H_{c2}(0)$  taken at $T = 120$~mK and open diamond for an extrapolation to 40~mK, for $H<H_{c2}(0)$ taken from Ref.~\citen{Bianchi2010}. Open circles indicate $\gamma (H)$ from the present experiments and the adjustment of $\Delta \gamma$ has been made at $H = 3$~T. Triangles: $\gamma_e (H)$ derived from resistivity experiments~\cite{Howald2011}. }
\label{f4_5}
\end{figure}

From the $T^2$ law of $M(T)$ measured below $H_{c2}(0)$, $\beta (H)$ is determined, then $\partial \gamma /\partial H$ (see Fig.\ref{dgammadH}) and by integration the relative field variation $\Delta \gamma (H_{c2}-\epsilon) \sim 1.1$~Jmole$^{-1}$K$^{-2}$ up to $H_{c2}-\epsilon =4.875$~T. $\Delta\gamma (H)$ is shown in the inset Fig.\ref{f4_5}(a). In the main panel Fig.\ref{f4_5}(b), the data have been offset by a constant, adjusted in comparison to direct specific heat data \cite{Bianchi2010} at $H = 3$~T, giving  $\gamma (H_{c2})\sim 1.2$~Jmole$^{-1}$K$^{-2}$. 
The field dependence of $\partial\gamma /\partial H$ does not show any distinct features that could be construed as the manifestation of a phase transition associated with the entrance into the $Q$ state as observed for example in the field dependence of $C/T$ for $H\parallel a$ at 250~mK in Ref.~\citen{An2010}. The absence of a $Q$-phase in CeCoIn$_5$ for $H \parallel c$ is in good agreement with recent neutron scattering experiments claiming that a misalignment of 17$^\circ$ from the $a$ axis to $c$ axis leads to a collapse of the signal of the $Q$-phase \cite{Blackburn2010}.
 However, we point out that a shoulder-like anomaly in the field dependence of $\partial \gamma / \partial H$ occurs in the vicinity of the vortex lattice phase transition at $H_{hex} \approx 4.4$~T. In addition, different experimental techniques have indicated other characteristic critical fields associated with the interplay between SC and AF. Thus, for example, the location of the respective fields $H_\alpha$ and $H_\rho$ recently derived by thermal expansion and resistivity experiments are shown in the inset of Fig.~\ref{dgammadH}\cite{Zaum2010, Howald2011}.

Above $H_{c2}$ the quasilinear $T$ dependence of $M(H)$ implies that the field dependence of $C/T$ is very weak. This conclusion is in good agreement with the weak field dependence of $C/T$ measured at $T = 250$~mK \cite{Ikeda2001} but does not agree with the field dependence of $C/T$ reported at $T \sim 120$~mK after the subtraction of the very large hyperfine nuclear specific heat coming from Co and In nuclei 
\cite{Bianchi2003b, Bianchi2002, Ronning2005}. Both elements  Co and In have isotopes with large dipolar and quadrupolar moments. Thus a nuclear contribution has to be taken into account for the specific heat and magnetization analysis. Assuming a sole nuclear Zeeman term of the form $A H^2/T^2$, this additional term to $\partial \gamma / \partial H$ gives a linear field correction $2AH/T^2$. The relative importance of the nuclear dependence will depend on the magnitude  of the other contributions to the $\partial \gamma / \partial H$ derivative. Above $H_{c2}$ these latter contributions are small at least above 100~mK by comparison to its value at $H_{c2}-\epsilon$, and the hyperfine contribution gives a significant channel for an increase in the $T$ dependence of $M$ on cooling leading to an apparent large temperature window where a linear temperature dependence of $M(T)$ is observed. Thus we cannot derive just above $H_{c2} (0)$ the variation of $\gamma (H)$. To estimate this value in the normal phase at $H_{c2}(0)$, attempts have been made through the extrapolation of $C/T$ to $T =0$ or through the indirect evaluation via the inelastic contribution to electron scattering. 

The precise determination of the $A(H)$ coefficient of the $T^2$ term of the resistivity \cite{Howald2011} (obeyed only below 40 mK) for $H > H_{c2}(0)$ confirms the quasi-divergence of $A(H)$ close to  $H_{c2}(0)$ \cite{Paglione2003}. In the physics of heavy fermion compounds the so-called Kadowaki-Woods relation $A \propto \gamma ^2$ is often obeyed due to strong local fluctuations. Scaling of $\sqrt {A(H)}$ with $\gamma (H)$ predicts a value of $\gamma_e (H_{c2} + \epsilon) \sim 2.25$~Jmole$^{-1}$K$^{-2}$ roughly in agreement with the low temperature extrapolation  of the electronic contribution of $C/T$ at 40~mK (see Fig.~\ref{f4_5}). A discontinuity of the $\gamma$ term at $H_{c2}$ seems to occur in association with the first order nature of the metamagnetic phenomena at $H_{c2} (0)$. 
Unfortunately,  dHvA signals taken while varying the field through $H_{c2}$ detect only the "light" $\alpha_3$ branch \cite{Settai2001}. However, the striking result is that the effective mass increases from $15 m_0$ in the field range from 4.85 to 4.95~T to $18 m_0$ for a scan from 5~T to 5.2~T.    A weak field dependence of $\gamma (H)$ appears for $H > 7$~T, the hypothetical value for a second order SC phase transition. Evidences that for $H > 7$~T another regime is achieved are also found in high field specific heat data.\cite{Kim2001}

The present results,  for $H<H_{c2}$, clearly demonstrated that $\gamma (H)$ reaches its singularity at $H_{c2}(0)$.
There is no evidence of a $Q$-phase at least characterized by a singularity in the field dependence of $\gamma$ below $H_{c2}(0)$, at least down to $(H_{c2}-H)/H_{c2}>0.02$.

\section*{Acknowledgment}
We thank A.~Bianchi and R. Movshovich for the communication of the specific heat data. The discussions with A.~Buzdin, K.~Miyake, V.~Mitrovic, T.~Sakakibara, Y.~Yanase, and G.~Varelogiannis have been very helpful. The work has been supported by the French ANR Delice, Sinus, and Cormat and the ERC project NewHeavyFermion. 


\end{document}